\documentclass[chicago]{emulateapj}
\usepackage{graphicx,amsmath,natbib,bm}
\usepackage[none]{hyphenat}
\usepackage{amsfonts}
\usepackage[top=1.3in, bottom=-0.3in, left=0.9in, right=0.35in]{geometry}
\usepackage{times}
\usepackage{enumitem}
\usepackage{framed}

\shortauthors{Y. Hezaveh et al.}

\begin{document}

\title{Probing the inner kpc of massive galaxies with strong gravitational lensing}
\author{Yashar D. Hezaveh, Philip J. Marshall, Roger D. Blandford}  
\affil{Kavli Institute for Particle Astrophysics and Cosmology, Stanford University, Stanford, CA, USA}

\begin{abstract} 

\noindent We examine the prospects of detecting demagnified images 
of gravitational lenses in observations of strongly lensed mm-wave 
molecular emission lines with ALMA. We model the lensing galaxies 
as a superposition of a dark matter component, a stellar component, 
and a central supermassive black hole and assess the detectability 
of the central images for a range of relevant parameters (e.g., 
stellar core, black hole mass, and source size). 
We find that over a large range 
of plausible parameters, future deep observations of lensed 
molecular lines with ALMA should enable detection of the central 
images at $\gtrsim 3\sigma$ significance. 
We use a Fisher analysis 
to examine the constraints that could be placed on these parameters 
in various scenarios and find that for large stellar cores, both 
the core size and the mass of the central SMBHs can be accurately 
measured. 
We also study the prospects for detecting binary SMBHs 
with such observations and find that only under rare conditions and 
with very long integrations ($\sim$40-hr) the masses of both SMBHs 
may be measured using the distortions of central images.

\end{abstract}


\section{introduction}

Probing the matter distribution in the innermost kpc of galaxies can answer key questions about super-massive black holes (SMBH), galaxy formation, and dark matter. It is now established that almost every massive galaxy harbors a SMBH at its center with a mass that strongly correlates with the mass of the host galaxy \citep{Kormendy:95,Ferrarese:00,Gebhardt:00,Tremaine:02}.
In addition, the distribution of stellar populations in central regions of galaxies contains information about their past merger histories and SMBH-stellar population interactions \citep[e.g.,][]{Barnes:92,Ebisuzaki:91,Barnabe:14}.  Various dark matter models also predict different structures for the central regions of dark matter halos \citep[e.g.,][]{Rocha:13}.  Mapping the matter density in the central regions of galaxies can thus shed light on various astrophysical phenomena.

Studies of local galaxies at optical wavelengths have shown that, unlike their lower-mass counterparts, the most massive elliptical galaxies often exhibit cored stellar light profiles, with core sizes ranging from 50 to 500 pc \citep[e.g.,][]{Ferrarese:06}. 
These galaxies are thought to form through gas-poor or ``dry'' mergers. In such mergers, the central structure of the resulting galaxy is dominated by the inner structure of the more concentrated progenitor.  Since high-mass ellipticals are thought to form from mergers of their lower-mass counterparts, with steep profiles, the existence of cores in these galaxies represents a challenge to our understanding of galaxy evolution. Cores in massive ellipticals, therefore, may be the result of a different process, such as  
``black hole scouring'' \citep{Thomas:14}.

It is thought that during a merger, the SMBHs of the two merging galaxies form a binary, which sinks to the center of the potential. The two orbiting SMBHs then dissipate angular momentum through three-body interactions with nearby central stars, pushing the stars to higher orbits and ``scouring out'' a core. This angular momentum loss may then allow the two black holes to merge \citep{Begelman:80}.
Previous studies have shown that the core sizes in these galaxies scale with the mass of their SMBHs, in agreement with theoretical predictions \citep{Kormendy:09,Kormendy:13}. It should be emphasized that these black holes are expected to be present and may have larger masses than in comparable nearby galaxies \citep{Bennert:10}. Direct measurements, however have been limited to low redshifts, since both dynamical measurements to constrain the stellar and SMBH masses, and morphological measurements to constrain core sizes require very high physical resolutions.

\begin{figure}
\begin{center}
\centering
\includegraphics[trim= 10 0 0 0, clip, width=0.48\textwidth]{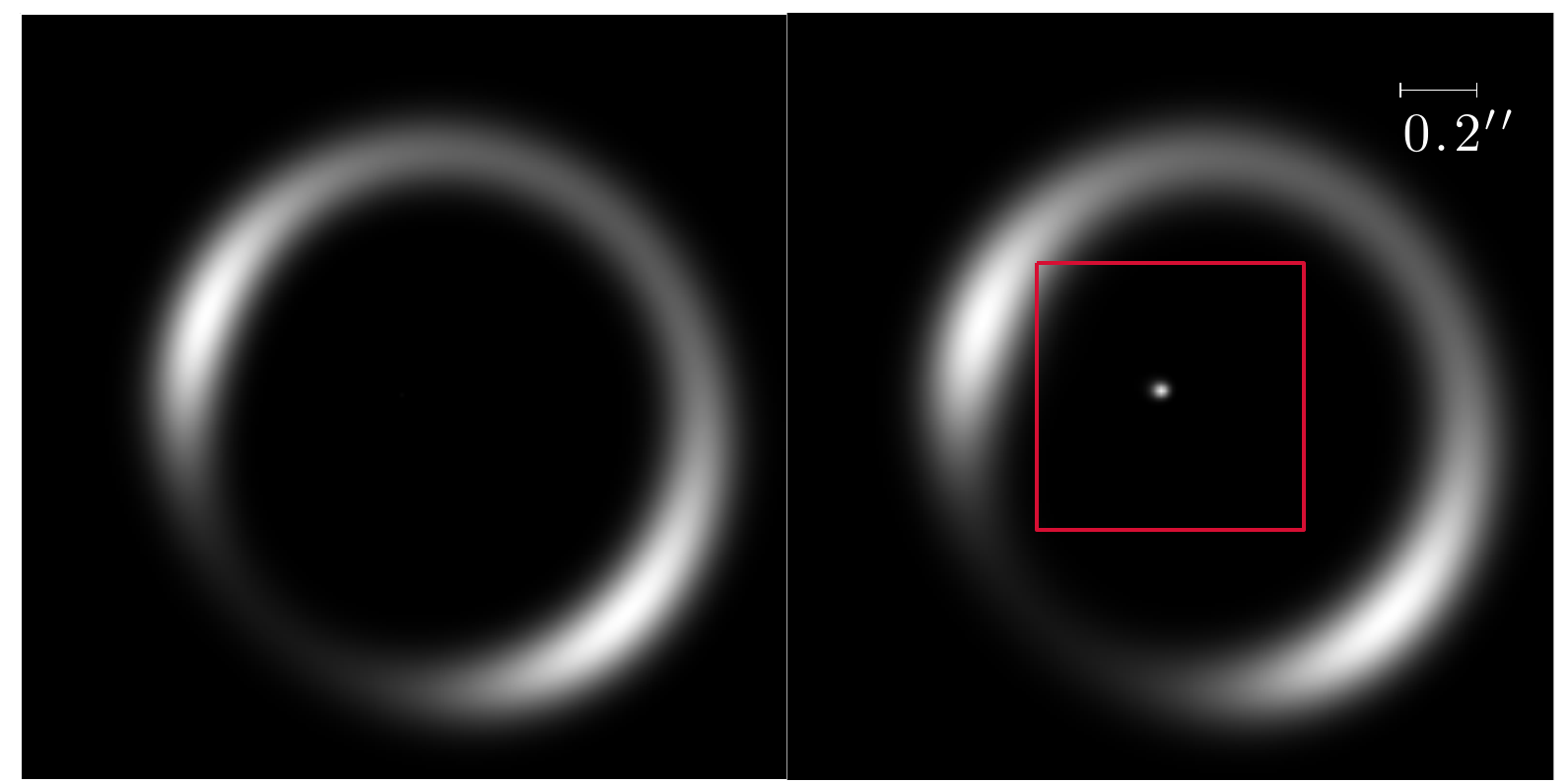}
\centering
\end{center}
\caption{ Left panel shows the strongly lensed images of a background source with a Gaussian light profile (FWHM 1170 pc), when the lensing potential is very steep and cuspy, resulting in large demagnification (and disappearance) of the central image.  The right panel shows images for the same lens parameters and a core of 400 pc. The red box is centered on the  central image.
\label{fig1}}
\end{figure}

Strong gravitational lensing is a powerful tool for probing the 
matter distribution in distant galaxies. Among other things, strong 
lenses have been used to constrain galaxy masses, density profiles, 
and abundance of dark matter subhalos \citep[e.g.,][]{dalal:02, 
Gavazzi:07, Bolton:08}. For double and quad image configurations, 
the theory of strong lensing predicts a third and a fifth image 
near the centers of lensing galaxies.
However, realistic galaxy density profiles produce central 
images that are almost always 
\emph{demagnified},\footnote{approximately as 
$(\Sigma/\Sigma_{crit})^{-2}$, where $\Sigma$ is the surface 
density and $\Sigma_{crit}$ is the critical surface density.} 
making their detection difficult \citep[e.g.,][]{Jackson:13}. If 
the lens galaxies have cuspy central profiles with steep slopes or 
very massive SMBHs, this central image may not form at all.
It is well-understood that the magnification of the central images 
is very sensitive to the matter distribution in the innermost 
regions of lens galaxies: very steep density profiles significantly 
demagnify the central images, whereas cored or shallow profiles 
render them brighter (Figure \ref{fig1}).

In addition to their low flux, the fact that central images coincide with the emission from lens galaxies makes their detection even harder. Distinguishing the central images from emission originating in the lens galaxies is extremely challenging.
If observed in the optical, absorption in the central dense regions of the lens can make the central images even dimmer, while the photon noise from the lens emission further reduces the sensitivity.

The individual behavior and statistical properties of central images in lensed populations have been extensively studied \citep[e.g.,][]{Wallington:93, Mao:01, Evans:02, Keeton:03}. These studies show that central images could have a wide range of magnifications \citep[e.g.,][]{Keeton:03}, and that cuspy profiles (or SMBHs) can suppress the flux of the central images \citep[e.g.,][]{Bowman:04,Rusin:05,Mao:12}.
Observational searches for these images have found a number of candidates. Although observations
of central images in group and cluster lenses are not uncommon \citep[e.g.,][]{Inada:08}, only one secure detection of a central image of a galaxy-scale lens exists to date \citep{Winn:04}. To avoid the possibility of absorption in the lens, most searches have focused on lensed radio sources. The radio spectra of candidate central images have been used to distinguish them from potential faint emission from the lens galaxies \citep[e.g.,][]{Zhang:07}.

A new large population of strong lenses has recently been discovered in mm-wave. These systems were initially detected as bright point sources in wide area mm/submm surveys \citep{vieira:10,negrello:10}. Follow-up observations have confirmed that they constitute a large population of strong lenses \citep{vieira:13, hezaveh:13b, bussmann:13}.
In particular, high resolution ALMA followup observations of these sources have revealed that the background galaxies are dusty, starburst, high redshift galaxies, containing a wealth of cold molecular gas \citep{Weiss:13}.  These observations showed that the extreme brightness of the sources, in combination with the high sensitivity of ALMA, result in very high signal to noise observations\footnote{For example, the lens models in \citet{hezaveh:13b} were based on just $\sim50$ second long observations.}.
Motivated by the discovery of this population and the operation of ALMA, here we revisit the issue of detecting central images.
Deep ALMA observations of molecular line emission in these sources are likely to be carried out for various reasons \citep[e.g.,][]{hezaveh:14a,hezaveh:14b}. If a central image of a lensed molecular line is detected in such observations, it will be readily identifiable, since it will correspond to the redshift of the source, leaving no doubt about its origin. In addition, since these lines are in mm-wave, the line fluxes are very unlikely to be suppressed by absorption in the lens. 
In contrast to lensed quasars, the sources are extended over hundreds of parsecs and suffer a range of demagnifications, further increasing the flux of the central images \citep[analogous to magnification damping of highly magnified images due to finite source effects, see e.g.,][]{hezaveh:11}.

 In this letter, we explore the possibility of detecting central lensed images in such deep observations and investigate the constraints that could be placed on the core size, mass of SMBHs, and the slope of the density profiles from detection, or non-detection of such images. We also examine if 
 such observations can possibly detect binary SMBHs, using the distortions they may cause in the central images. 
 In Section 2 we describe the simulations, in Section 3 we present the results and discuss them, and finally conclude in Section 4. We use a flat $\Lambda$CDM cosmology with $h=0.71$ and $\Omega_M=0.267$.

\begin{figure}
\begin{center}
\centering
\includegraphics[trim= 0 0 5 6, clip, width=0.48\textwidth]{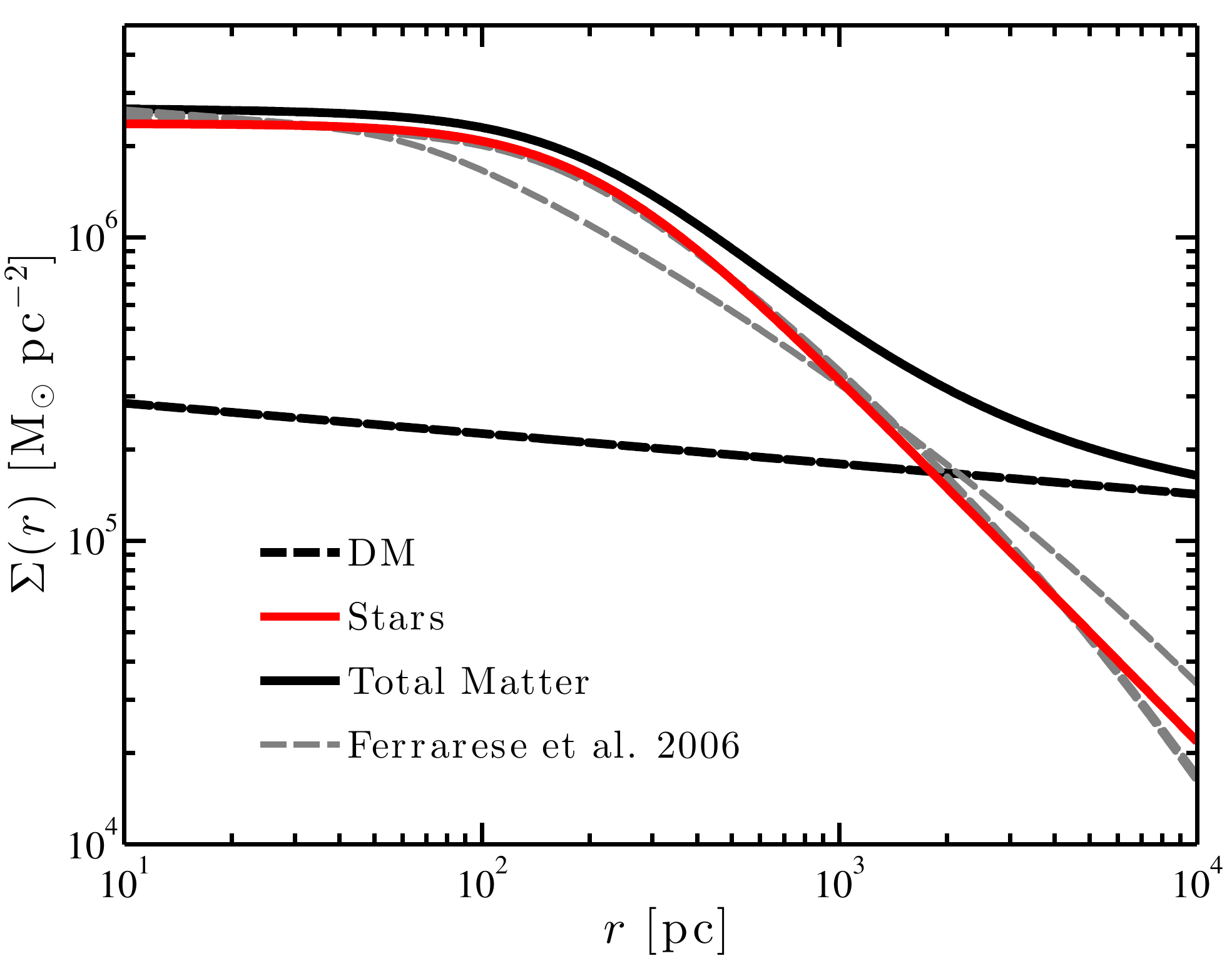}
\centering
\end{center}
\vspace{-1mm}
\caption{ Model density profiles. The black dashed line shows the dark matter component, with a slope of 0.1, consistent with a projected NFW at the innermost regions. the grey dashed curves show cored-sersic five fits to stellar light profiles of massive galaxies from Ferrarese et al. (2006). The red solid curve shows our cored-power-law model that we use to approximate this stellar component. The solid black line shows the total matter density (dark plus stellar) of our model. 
\label{fig2}}
\vspace{-1.5mm}
\end{figure}

\section{Simulations}
We generated lensed images of background sources, predicted ALMA visibilities, and used them to estimate the detection significance of central images for various parameters. We simulated observations of a high-$J$ CO line. The line is assumed to have a velocity integrated flux of 1 Jy km/s and a FWHM of 400 km/s, resulting in an average flux of 2.5 mJy over a 400 km/s band, a typical value for high redshift, dusty galaxies \citep{Bothwell:12}. 

We modeled the lens potential as a sum of three components: dark matter, stellar population and a single central SMBH. The dark matter component was modeled as an elliptical singular power-law with a slope of 0.1, in agreement with the projected surface mass density of the NFW profile in the innermost regions \citep{Golse:02}. As  pointed out by \citet{Keeton:03}, due to its expected flatness in these regimes, the dark matter component has negligible influence on the central images (although models with dark matter cusps have been entertained). 
The stellar population is modeled as an elliptical cored power-law, $\Sigma(r)\propto (r^2+R_{\mathrm{core}}^2)^{\gamma_s}$. The deflection angles for elliptical cored power-law models were calculated using FASTELL \citep{Barkana:98}.  Figure \ref{fig2} shows the stellar component (red curve), dark matter component (black dashed curve) and the sum of the two (black solid curve). The grey dashed curves show a few examples of core-sersic models with parameters taken from \citet{Ferrarese:06}. Although the cored power-law model used in this work does not account for the slope of the stellar distribution below the core break, it is a close fit and a reasonable model to approximate the observed stellar light profiles in the inner few kpc. 
The black hole was modeled as a simple point mass at the center of the potential. 
Lensing galaxies are typically massive ellipticals with velocity dispersions of order 200 km/s \citep[e.g.,][]{Sonnenfeld:13}. From the M-$\sigma$ relation, we predict that these galaxies should contain black holes with masses in excess of $2\times10^8 M_{\odot}$ \citep[e.g.,][]{Kormendy:13}. In our simulations we have chosen black holes with masses of 1, 2, and $4 \times 10^8 M_{\odot}$.

We assumed that all the three components are concentric, though we note that two- and three-black hole interactions may invalidate this assumption.
The stellar and dark mass of the lens were normalized such that they each contain $1.5\times10^{11} M_{\odot}$ in a radius of 10 kpc. The resulting Einstein radius is of order 1 arcsec, in agreement with galaxy-galaxy strong lenses \citep[e.g.,][]{hezaveh:13b}.
To predict the visibilities, we calculated the ALMA $uv$-coverage of a 10-$hr$ long observation (full array), using the $simobserve$ task of Common Astronomy Software Applications package, for an observing frequency of 150 GHz. 
The model visibilities were calculated by computing the 2D Fourier transform of the surface brightness data maps and resampling the Fourier transform maps over the $uv$-coverage. 
The noise was estimated using ALMA sensitivity calculator for a channel width of 400 km/s at 150 GHz.

For simulations where we calculated the detection significance of the demagnified image, we computed the magnification at every pixel in the image plane and generated maps with and without the demagnified flux and evaluated the detection significance of the central image by comparing the two resulting visibility sets. In other simulations, where the constrains on parameters were needed, we used a Fisher analysis to compute the full covariance of all parameters. 
We used finite differencing of visibilities to calculate the gradients, and marginalize over the nuisance parameters (e.g., source position, lens ellipticity), assuming uniform priors, to calculate the posterior of the relevant parameters.

\begin{figure}
\begin{center}
\centering
\includegraphics[trim= 0 0 0 0, clip,width=0.48\textwidth]{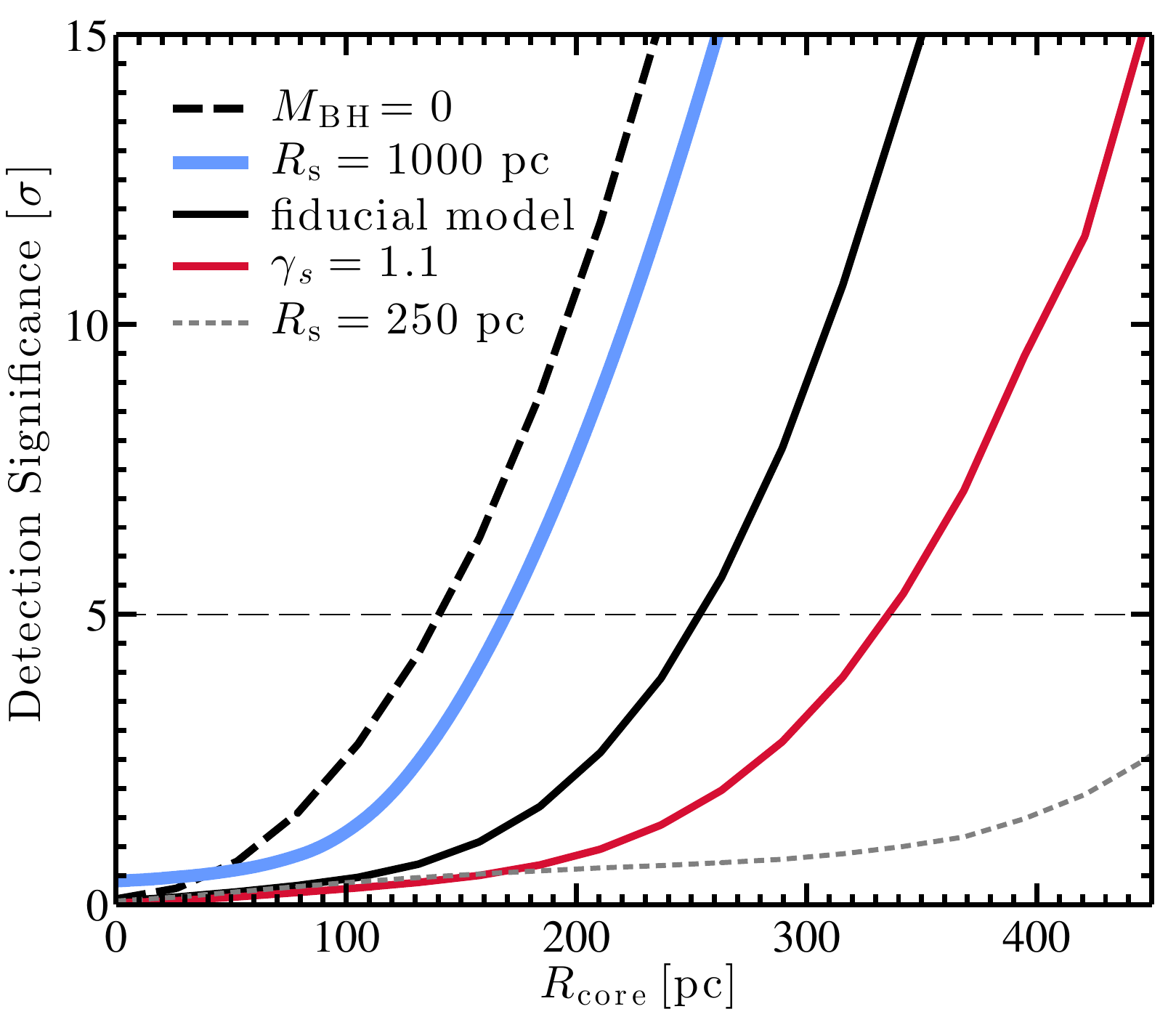}
\centering
\end{center}
\caption{ Detection significance of a central image, as a function of the stellar core size for a 10-hr long ALMA observation. 
The fiducial model (solid black curve) corresponds to $\gamma_{stellar}=1.0$, $M_{\mathrm{BH}}=2\times10^8 M_{\odot}$, and a source size of 500 pc (radius). The lens has an ellipticity of 0.2. The dashed black curve shows the detection significance when the SMBH is removed and the solid red curve shows the effect of a steeper density profile ($\gamma_{\mathrm{stellar}}=1.1$). The solid blue and gray dotted curves show the detections for a source with a radius of 1 kpc and 250 pc respectively. 
\label{fig:3}}
\end{figure}

\section{Results and Discussions}
\subsection{Detection of central images with ALMA}
The fluxes of central images depend principally on the slope of the lensing potential, the core size, and the mass of the central SMBHs.  Figure \ref{fig:3} shows the detection significance of a central image for a range of these parameters in a 10-hr long ALMA observation. 
The figure shows the detection significance of the demagnified central flux, as a function of the stellar core size.
The fiducial model (black curve) has a stellar profile slope of $\gamma_s=1.0$, and a $2\times10^8M_{\odot}$ SMBH, with a background source 
with a Gaussian profile with an rms of 500 pc.  The red curve shows the effect of a steeper density slope, suppressing the flux of the central image.

As seen in this figure, larger stellar cores result in brighter images. 
In the absence of a massive black hole, when $\gamma_s=1.0$ and with a core size larger than 150 pc, such observations should be able to detect the central images with more than $5\sigma$ significance. For significantly shallower ($\gamma_s<0.8$) profiles and larger cores, the central image is \emph{magnified} and its detection should be trivial.  The magnification and flux of central images is also influenced by SMBHs.
SMBHs demagnify the images on scales comparable to the size of their Einstein radii ($\sim30$ mas for a $2\times10^8M_{\odot}$ SMBH).
A comparison of the thick black dashed curve to the solid black curve in Figure \ref{fig:3} shows the suppression of the fluxes of central images in presence of a $2\times10^8M_{\odot}$ SMBH.

As seen in Figure \ref{fig:3}, and as pointed out by \citet{Keeton:03}, plausible parameters predict  a wide range of magnifications for   central images. Although for some parameter combinations (e.g., $\gamma_{s}=1.1,$ $M_{\mathrm{BH}} = 2\times10^8 M_{\odot}$,$R_{\mathrm{core}}<200$ pc ) there is little chance of detecting the central images, they may be detectable over a non-negligible fraction of the parameter space.

\begin{figure}
\begin{center}
\centering
\includegraphics[trim= 0 0 0 0, clip, width=0.48\textwidth]{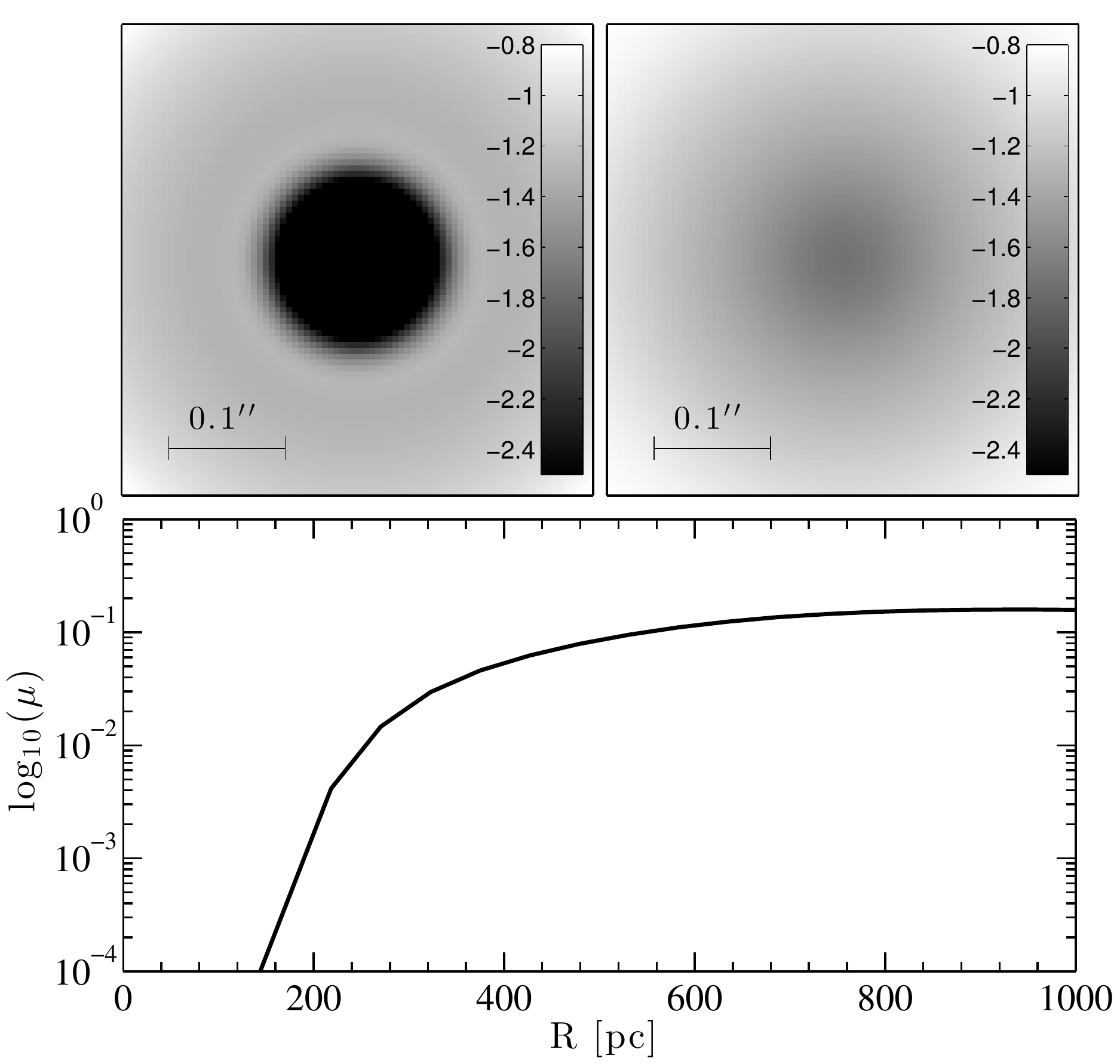}
\centering
\end{center}
\vspace{-1mm}
\caption{The two top panels show the source plane magnification 
maps for two sources, with different sizes (left: 100 pc, 
right 500 pc in radius) behind a lens containing a black hole with 
a mass of $1\times 10^8 M_{\odot}$. The magnification at the 
central regions of the left panel approaches zero and the central 
image does not form (the color scale has been clipped at 
$10^{-2.5}$). The lower panel shows the magnification at the center 
(when the source is perfectly behind the lens) as a function of 
source size).
\label{figMu}}
\vspace{-1.5mm}
\end{figure}

We also find that the detection significance of the central images 
can significantly \emph{increase} for an extended source with a 
radius of 1 kpc compared to a point source. Since extended sources 
occupy a larger area of the source plane, which may cover regions 
with higher magnifications, the resulting flux can be larger than 
the flux of more compact sources.

The solid blue curve in Figure \ref{fig:3} shows the increase in 
the flux of the demagnified image for a larger source (with a 
radius of 1 kpc). In contrast, the flux of a source with a radius 
of 250 pc can be severely supressed (dotted grey curve). This 
effect is illustrated in Figure \ref{figMu} where the effects of 
source size on magnification are shown. The two top panels of 
Figure~\ref{figMu} show the source plane magnification (flux 
magnification as a function of source position behind the lens) for 
two sources with different sizes (left: 100 pc, right: 500 pc in 
radius). For a compact source (top left panel) in the central 
regions the magnification approaches zero and the central image 
does not form. For more extended sources, however, even when the 
source is exactly behind the lens, the large extent of the source 
allows covering higher magnification regions, increasing the total 
flux magnification.
The bottom panel shows the magnification as a function of source 
size, for a source that is located on the lens axis.
As the source galaxies are undergoing prodigious star formation, 
these large source sizes are reasonable and in agreement with 
observations \citep{hezaveh:13b}.
More extended and intrinsically brighter emission lines such as 
C[II], the brightest molecular line in high redshift galaxies, may 
therefore be the most ideal targets for such observations.

\begin{figure}
\begin{center}
\centering
\includegraphics[trim= 2 5 0 0, clip, width=0.48\textwidth]{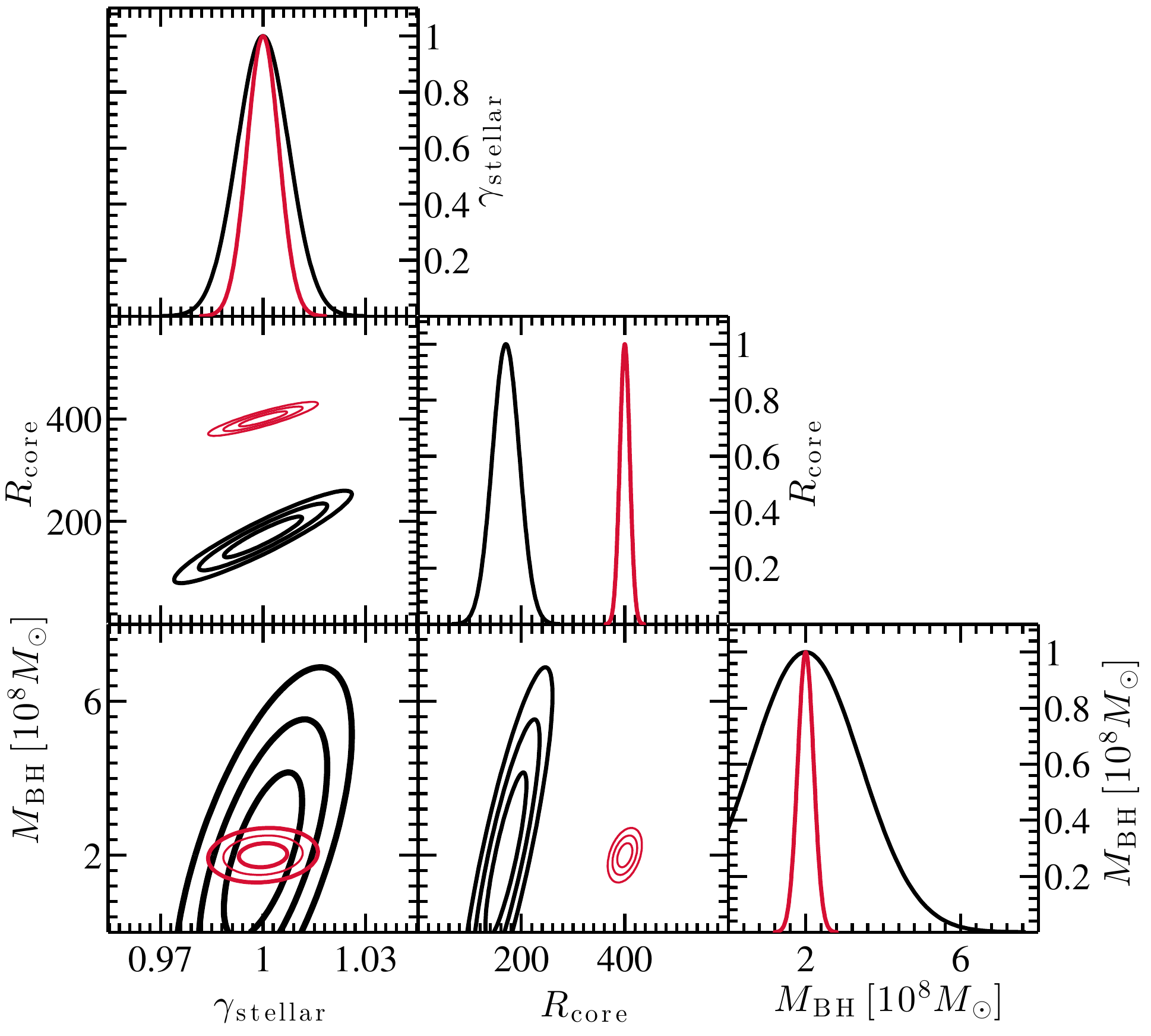}
\centering
\end{center}
\caption{ Parameter covariance matrices for two simulations (after marginalization over nuisance parameters). Black curves show that when the core size is small, the mass of the SMBH is not measured with high precision and only an upper limit can be placed on it. The bottom panel, shows that for larger core sizes, the mass of the SMBH could be measured with high significance. 
\label{fig:4}}
\end{figure}

\subsection{SMBHs and the core size}
Here, we study the constraints that could be placed on the slope of the density profile, the size of the stellar core, and the mass of the SMBHs in such deep observations. 
We performed a Fisher analysis to examine the constraints and parameter degeneracies in simulations in which the central images were detected, as well as those in which they were below the detection limit. All model parameters were included in the Fisher matrix, and nuisance parameters were later marginalized over (e.g. intrinsic source flux and position).  
 Figure \ref{fig:4} shows the parameter covariance for two simulations with different core radii (black contours:  $R_{\mathrm{core}}$=170 pc, red contours: $R_{\mathrm{core}}$=400 pc).
We found that when $\gamma_s=1$, core sizes larger than 100 pc can be measured with high significance. As seen in Figure \ref{fig:4}, 
for small cores, the mass of SMBH is highly degenerate with the core size: larger cores result in brighter images, whose flux can be supressed by a more massive SMBH.
This degeneracy,  however,  is reduced when the core size is very large: large cores produce more extended central images that are spatially resolved. The distortion caused by a SMBH can distort this image on smaller scales, resulting in a distinct, resolved \emph{dip} in its surface brightness. 
We find that for cores with a radius larger than $200$ pc, central SMBH masses can be measured with high significance (e.g., red contours in Figure \ref{fig:4}), but for smaller core sizes, only upper limits can be placed on the mass of SMBHs.
We also find that when the core size is larger, the constraints on the density profile slope are stronger.

\begin{figure}
\begin{center}
\centering
\includegraphics[trim= 20 0 20 0, width=0.46\textwidth]{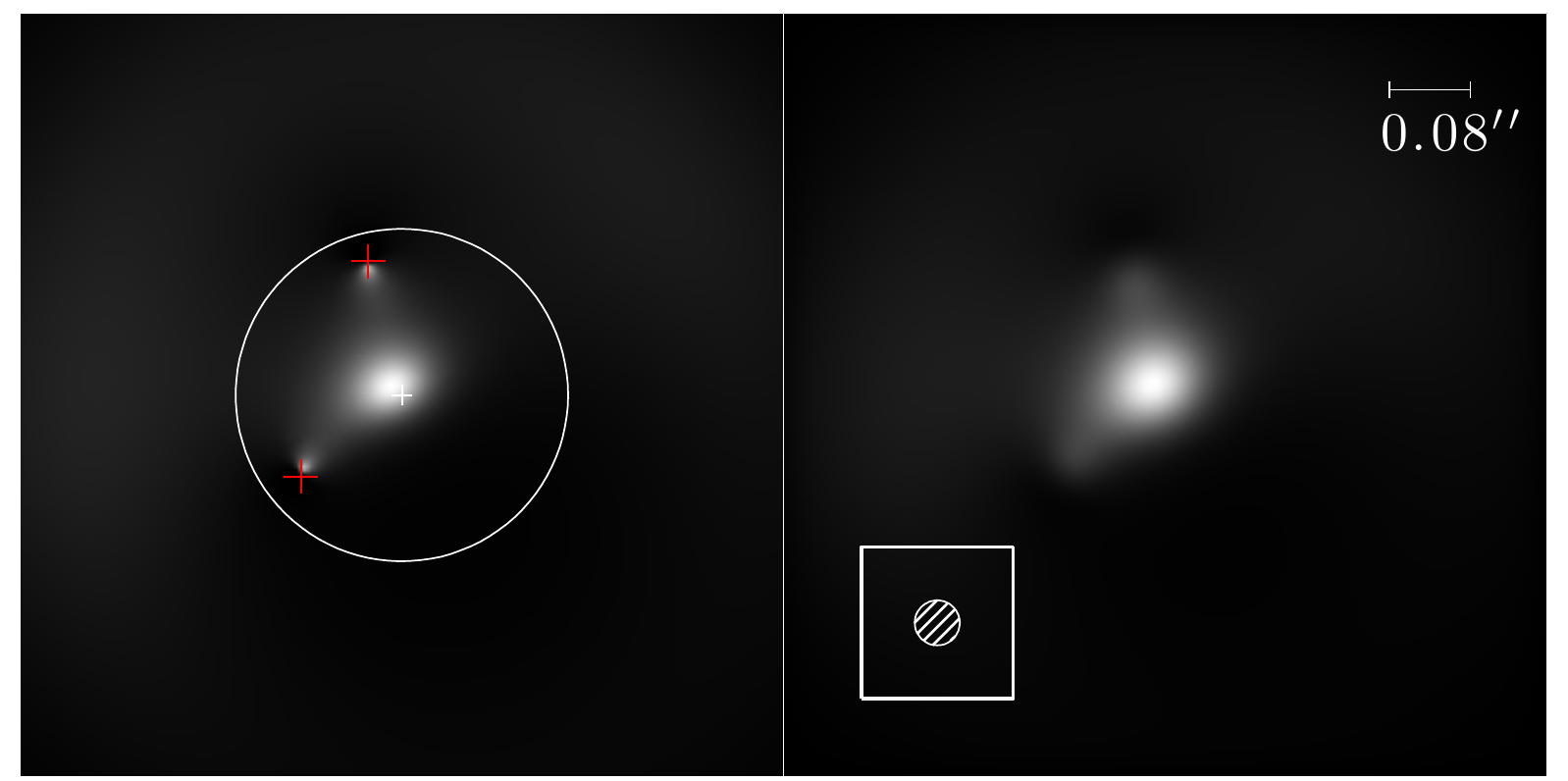}
\centering
\end{center}
\caption{ Lensing of a central image by two SMBHs with masses of $4\times 10^8M_{\odot}$. The left panel shows the simulated central image. The red crosses show the positions of the SMBHs. The white circle has a radius of 1 kpc  from the center of the galaxy. The right panel shows the same data after convolution with the ALMA beam. The beam is shown in the lower panel.
\label{fig:5}}
\end{figure}

\subsection{Detection of Binary SMBHs}
We also investigated the possibility of detecting binary SMBHs using the perturbations that they induce on lensed central images. 
In principle, several black holes could be held up in orbits in these young massive galaxies.
We placed two SMBHs with masses of $4\times10^8 M_{\odot}$ each in random positions inside a circle of radius of 1 kpc from the center of the galaxy and, using the Fisher analysis, computed the detection significance of their masses. 
We found that in most cases, it is not possible to measure the mass of both SMBHs, either because they are not near the central images or due to the degeneracy between their masses, their positions, and the core radius. 
However, when the central images are fairly extended (due to a shallow density slope and a large core size), a larger fraction of the inner 1 kpc region is covered by the central image, increasing the probability of lensing by SMBHs. Such images also have larger fluxes, increasing the signal to noise of the observations and they also allow for the detection of fairly separated SMBHs. In this case, the SMBHs can affect different parts of the central image, reducing the degeneracy between their masses.
Under such conditions, we found configurations that allowed a detection of both SMBH masses. The constraints on SMBH masses in such observations arise from the distortions in the resolved \emph{shape} of the central images. Figure \ref{fig:5} shows an example of such configuration when $\gamma_s = 0.9$ and $R_{\mathrm{core}}=300$ pc. Left panel shows the central image with the positions of SMBHs marked with red crosses. The distortion in the central image by the SMBHs is clearly visible. The right panel shows the same image after convolution with ALMA beam. Both SMBHs are detected at $\sim5\sigma$.

To estimate the probability of configurations which allow detection of binary SMBHs, we simulated many realizations of the positions of the SMBHs for a high resolution 40-h long observation. 
We found that $\sim10\%$ of the simulations, in which the core size was larger than 200 pc, both SMBHs were detected. 
This number is much larger than what is calculated in \citet{Li:12}, since they estimated  the probabilities for point sources, appropriate for lensed quasar samples. The extended structure of the background sources studied here can significantly increase the probability of overlapping the black hole caustics, increasing the lensing cross-section of SMBHs.

  Although this number may seems promising, the prospects of such measurements will be limited by the number of discovered bright central images, which based on their current dearth, may be few. We conclude that, only under rare conditions, and with very long observations ($>40$ hours), central image may allow us to probe binary SMBHs. The prospects of such measurement will ultimately depend on discoveries of much larger samples of lenses and the dynamical evolution of the black holes.

\section{Conclusions}
We have examined the prospects of detecting central, demagnified images of strongly lensed molecular lines with ALMA. We found that there is a non-negligible space of realistic and plausible parameters that result in a high significance detection of the central images. We showed that such deep observations can either allow a measurement of the mass of the central SMBH and the stellar core size, or place strong limits on them.
We also studied the possibility of detecting binary SMBHs and found that only under very rare conditions central images may allow detection of both SMBHs. Successful execution of such programs could teach us much about evolution of galactic nuclei.

\acknowledgements{
We thank Risa Wechsler, Matt Becker, and the anonymous referee for useful discussions and helpful suggestions.
YH acknowledges support from the NSF grant 0807458.
The work of PJM was supported in part by the U.S.
Department of Energy under contract number DE-AC02-76SF00515.
}

\bibliographystyle{apj}

\end{document}